\documentstyle[prl,aps,multicol]{revtex}
\begin{document}
\draft

\begin{multicols}{2}

\noindent
{\bf Comment on ``Critical Behavior in Disordered Quantum Systems
Modified by Broken Time--Reversal Symmetry''} 

    In a recent Letter\cite{HP} Hussein and Pato (HP) have employed
    the maximum entropy principle (MEP) in order to derive
    interpolating ensembles between any pair of universality classes
    in random matrix theory (RMT).  In their procedure the entropy of
    the distribution function over the matrix ensemble
    $S=-\int\,dH\,P(H)\,\ln\,P(H)$ is maximized subject to the
    constraints $\langle 1\rangle=1$ and $\langle$Tr$(H^2)\rangle
    =\mu$ and moreover $\langle$Tr$[(H-H_0)^2]\rangle =\nu$. In the
    latter condition $H_0$ is assumed to be defined in the subspace of
    $H$. The symmetry of $H$ is incorporated in $\mu$ and parameter
    $\nu$ is used to drive the ensemble to another universality class.
    The authors have worked out the transition from GUE to GOE in
    detail and admit that their results are equivalent to the
    convolution construction by\cite{LH}. Subsequently the authors
    apply their method for the transition from an ensemble of the RMT
    to the Poissonian ensemble (PE), where $H_0$ is just the diagonal
    of $H$. The matrix elements of $H$ are Gaussian distributed with
    distinct variances for diagonal (denoted as $1/2\alpha_0$) and
    off-diagonal ($1/2(\alpha_0+\beta)$) matrix elements. In what
    follows HP numerically calculate the average entropy of the
    eigenstates for fixed matrix size $N$ and believe to find the
    ensemble corresponding to the critical point in a metal--insulator
    transition (MIT) with the help of its maximal derivative as a
    function of $\beta/\alpha_0$.

    In the present Comment we focus our attention to the part in
    connection with the MIT in\cite{HP}. First of all it is
    straightforward to show that for the case of the RMT$\to$PE
    transition the model of \cite{HP} is equivalent to a model
    investigated recently\cite{Boris,G-LS-PS} which has its roots in
    an early paper \cite{RP}. It has been interesting to learn from
    \cite{HP} that in fact the ensembles studied
    in\cite{Boris,G-LS-PS,RP} can all be constructed using the MEP.

    However, the application of these matrix ensembles to the problem
    of the MIT suffers from several deficiencies that are already
    pointed out in most of the above mentioned works.

    1. The search for a critical ensemble for fixed $N$ is an improper
    choice, one should rather address this question in the
    $N\to\infty$ limit as in\cite{Boris}. We have performed similar
    numerical simulations as \cite{HP} and have seen very strong
    $N$-dependence in Fig. (1) of \cite{HP}. It manifests in two ways.
    Both the transition range is shifted and the maximal average
    entropy corresponding to the fully chaotic eigenstates depends on
    $N$, as well. Hardly any surprise for the latter, since the
    entropy is an extensive quantity and should be monotonously
    increasing with $N$. Note that the entropy of the eigenstates
    still serves as an appropriate tool for the characterization of
    the states changing their character from extended to localized
    \cite{PV}.

    2. In the $N\to \infty$ limit of the HP model the following
    differences to the MIT occur.

    ($i$) For noninteracting electrons the MIT occurs only in
    dimension $d>2$ and critical exponents are known to depend on $d$.
    The HP model, however, contains no spatial dimensionality.

    ($ii$) Furthermore, the HP model doesn't exhibit a mobility edge
    in the usual sense. There is no transition from localized to
    extended states when the energy within the band is changed.

    ($iii$) Most seriously, in the $N\to\infty$ limit, all states in
    the band are either localized or extended, depending on how the
    ratio $(\alpha_0+\beta)/\alpha_0$ scales with $N$\cite{Boris}. A
    new ensemble, intermediate between RMT and PE, occurs (in the
    $N\to\infty$ limit) when this ratio scales as $N^2$. The extended
    states of these critical ensembles are still qualitatively
    distinct from those of the MIT since they are not
    multifractal\cite{Martin}. This can be manifested e.g. in the
    level compressibility and in the energy level spacing
    distribution, $P(s)$, for large level--separations. According to
    recent numerical simulations at the MIT the form is a simple
    exponential\cite{Num} $P(s)\propto\exp(-1.9s)$ in $d=3$, in
    contrast to Eq.~(14) of \cite{HP}.

    Finally we mention that new matrix models have been considered
    recently \cite{KM} that can account for a transition and
    multifractality of the critical eigenstates \cite{Martin}.

    This work has been partially financed by DFG, SFB-341, and OTKA
    Nos. T021228, T024136, F024135).

\medskip\noindent
    Martin Janssen$^a$, Boris Shapiro$^b$, and Imre
    Varga$^{a,c}$\medskip\\
    {\small $^a$Institut f\"ur Theoretische Physik, Universit\"at zu
    K\"oln\\
    \hphantom{$^a$}Z\"ulpicher Str. 77, 50937 K\"oln, Germany\\
    $^b$Department of Physics, Technion, 32000 Haifa, Israel\\
    $^c$Department of Theoretical Physics, Technical University of\\
    \hphantom{$^c$}Budapest, H-1521 Budapest, Hungary

\vskip0.2cm\noindent
    Received \today}  

\end{multicols}
\end{document}